\documentstyle[prl,aps]{revtex}
\draft
\begin{document}

\title{ Note on the luminosity distance }
\author{  Edward Malec,  Grzegorz Wyl\c e\.zek and Janusz Karkowski  }
\address{  Institute of Physics,
Jagiellonian University,
30-064 Krak\'ow, Reymonta 4, Poland   }

\maketitle

\begin{abstract}
We re-derive  a formula relating the areal and luminosity distances,
entirely in the framework of the classical Maxwell theory, assuming
a geometric-optics type condition.

\end{abstract}

\pacs{  }
\date{\today }

\section{ Introduction}

It is accepted since Robertson \cite{Robertson} that the luminosity distance
relates to  the areal distance \cite{1} as follows, in   Friedman-Lemaitre-Robertson-Walker
(FLWR henceforth) spacetimes,
\begin{equation}
D=   (1+z) R.
\label{0}
\end{equation}
Formula (\ref{0}) is usually derived in a way that mixes quantum
mechanical  and classical concepts \cite{Weinberg}. Robertson has shown
this result  within the framework of the classical Maxwell theory using
conservation laws.  His approach suffers from a number of minor erronous
statements and gaps in the reasoning.  One of that   concerns   the form of
the electromagnetic energy-momentum tensor, another
the   energy flux. A   comment on the latter.
The Maxwell equations in FLRW
spacetimes  possess  a mathematical strictly conserved
energy which is different -- as it is quite common in curved spacetimes
  -- from the 'proper energy', the total energy
detected by a {\it comoving observer}. The latter is directly
related to the luminosity, but is not conserved; in fact it is equal to
the former energy divided by the conformal factor (see below).
Robertson   assumes that the corresponding energy fluxes
 scale in the same way as the energies themselves; that can be true only
approximately (but including all  cases of  astrophysical interest), as
will be shown.

These shortcomings do not  influence the validity of the main conclusion of
\cite{Robertson}. The  aim of this paper is to provide a  derivation for the
cosmological distance formula which is consistently classical and fills some
gaps in the proof of Robertson \cite{Robertson}. The order of the rest of
this paper is following. Sec. 2 brings fundamental definitions and equations.
Sec. 3 is dedicated to the derivation of an explicit solution of
the Maxwell equations. Next section shows how one incorporates initial data into
this closed form solution.  The first part of Sec. 5 contains a proof
that the electromagnetic energy momentum tensor can satisfy  assumptions made
by Robertson. Sec 5B accomplishes the  proof of (\ref{0}).

\section{Equations}

The space-time geometry  is defined  by the FLWR  line element,

\begin{equation}
ds^2 = a(\eta)^2( - d\eta^2 + dr^2 + \sigma^2 d\Omega^2~),\label{1}
\end{equation}
where $\eta $ is a (conformal) time coordinate, $r$ is a radial coordinate
that coincides with the areal radius,
$d\Omega^2 = d\theta^2 + \sin^2\theta d\phi^2$ is the line element on
the unit sphere
with $0\le \phi < 2\pi $ and  $0\le \theta \le \pi $.  $a(\eta )$ is a conformal
factor and $\sigma $ is given by $\sinh{(r)}$ (for negative curvature $k=-1$),
$r$ (for flat spacetime $k=0$) and $\sin{(r)}$ (for positive curvature $k=1$ ).
 Throughout this paper $c$ and $G$,  the velocity of light and the
 gravitational coupling constant respectively, are put equal to 1.
 The standard cosmological time is related to $\eta $ by $ad\eta =dt$.

The Maxwell equations read
\begin{equation}
\nabla_{\mu }F^{\mu \nu }=0,\label{3}
\end{equation}
where $F_{\mu \nu }=\partial_{\mu }A_{\nu }- \partial_{\nu }A_{\mu }$ and
$A_{\mu }$ is  the electromagnetic  potential.
It is convenient to assume $A_0=0$ and the Coulomb gauge condition
$\nabla_iA^i=0$.  In such a case there are two independent degrees of
freedom,  represented by the  magnetic or electric  modes.

The  magnetic  modes are defined as follows
\begin{eqnarray}&& A=\Sigma_{l=1}^{\infty }\Sigma_{m=-l}^l{f_{lm} (\eta, r)
\over \sqrt{l(l+1)}}
\Biggl(  -m{Y_{lm}\over \sin \theta }d\theta\nonumber \\
&& -i\sin \theta \partial_{\theta }Y_{lm} d{\phi}  \Biggr) .
\label{4}
\end{eqnarray}
Here $Y_{lm}$ are the spherical harmonics. The multipole expansion coefficients
$f_{im}$ depend on the conformal time $\eta $ and the radial coordinate $r$.
Straightforward caclulation shows that the equations of motion reduce to
a system of hyperbolic equations
\begin{equation}
(-\partial_{\eta }^2 + \partial_{r }^2)f_{lm} = { l(l+1)\over \sigma^2}f_{lm}
\label{5}
\end{equation}
for the multipoles $f_{lm}(\eta, r)$.  

The   electric  potential  read
\begin{eqnarray}
&& A= \Sigma_{l=1}^{\infty }\Sigma_{m=-l}^l {\frac{1}{\sqrt{l(l+1) }}}
\Biggl(  {\sqrt{l(l+1) }}{h_{lm} (\eta, r)}{Y_{lm}}dr
  + \nonumber \\
&&+ {k_{lm} (\eta, r)}\partial_{\theta }Y_{lm}d{\theta } + \nonumber \\
&&+{k_{lm} (\eta, r)}im{Y_{lm}}d{\phi}  \Biggr)  .
\label{6}
\end{eqnarray}
Here  ${h_{lm} (\eta, r)}$ and ${k_{lm} (\eta, r)}$
are  multipoles. From the Coulomb gauge condition $\nabla_iA^i=0$
 we have the relation between   multipoles
${k_{lm}(\eta,r)}$ and ${h_{lm}(\eta,r)}$:
\begin{equation}
\sqrt{l(l+1)}{k_{lm}} = \partial_{r} ( \tilde{h}_{lm} )
\label{7}
\end{equation}
where $\tilde{h}_{lm} = {\sigma^2}{h}_{lm}$.
One  can show that  the first Maxwell equation $(\nabla_{\mu }F^{\mu 0 }=0)$
is identity. The second one,   $(\nabla_{\mu }F^{\mu r }=0)$,
reduces to the same hyperbolic equations  as in  Eq.(\ref{5}):
\begin{equation}
(-\partial_{\eta }^2 + \partial_{r }^2) \tilde{h}_{lm} = { l(l+1)\over
\sigma^2} \tilde{h}_{lm}.
\label{8}
\end{equation}
The   two remaining Maxwell equations $(\nabla_{\mu }F^{\mu k }=0$ where
$k = {\theta}, {\phi})$ yield
\begin{equation}
(-\partial_{\eta }^2 + \partial_{r }^2){k}_{lm} = { l(l+1)\over \sigma^2}
{k}_{lm} -  \frac{2 \sqrt{l(l+1)} (\partial_{r}{\sigma})}
{\sigma}{k}_{lm}.
\label{9}
\end{equation}
It's easy to show that (\ref{8}) is equivalent to (\ref{9}). In order to see
this,   differentiate both sides of (\ref{8}) with respect  $r$ and use the
Columb gauge condition (\ref{7}); a quick calculation gives (\ref{9}).

\section{Compact form of solutions}

Equations (\ref{5}) and {\ref{8}) have identical form; they can be written
compactly as
\begin{equation}
(-\partial_{\eta }^2 + \partial_{r }^2) \phi_{lm} = { l(l+1)\over \sigma^2}
\phi_{lm},
\label{9a}
\end{equation}
where $\phi=f, \tilde{h}$. Below we will drop the irrelevant index $m$.
The forthcoming constructions apply equally well
to magnetic and electric degrees of freedom. The form of Eq. (\ref{9a})
suggests that   each of the three  cosmological models should be discussed
separately, but we show later, that there exists a unique approach that
works in  all these cases. In this section we construct a solution of
Eq.(\ref{9a}).

Define a set of {\it generating functions} (as we  will sometimes refer to
them in the forthcoming text)  of the  form
\begin{equation}
f({\eta} , r)=f(r-\eta)
\label{10}
\end{equation}
and also
\begin{equation}
g({\eta} , r)=g(r+\eta).
\label{11}
\end{equation}
{\bf Lemma.} For any $f$ and $g$ the     function
\begin{equation}
\phi_{l}(r,\eta) = \sigma^l\,\underbrace{ {\partial_{r}}\,{\frac{1}
{\sigma}}\,{\partial_{r}}\,{\frac{1}{\sigma}}\cdots
{\partial_{r}}\,\bigl(\frac{f+g}{\sigma}}_{l}\bigr)
\label{12}
\end{equation}
 solves  Eq.  (\ref{9a}).

{\bf Proof of Lemma.}

The proof uses the method of   matematical induction.  One can write the
right hand side of (\ref{12})  as $ \sigma^l \partial_{r}\Biggl( \Bigl(
{ \sigma^{l-1} \over \sigma^{l-1}} \Bigr)
 \underbrace{  \,{\frac{1}{\sigma}}\cdots
{\partial_{r}}\,\bigl(\frac{f+g}{\sigma}}_{l-1}\bigr)\Biggr) $  (herein we
simply put  ${ \sigma^{l-1} \over \sigma^{l-1}}$ after the first  sign of
differentiation). Thence one obtains a recursive formula
\begin{equation}
\phi_{l}(r,\eta) = \partial_{r}\phi_{(l-1)}(r,\eta)-l{\frac{\partial_{r}
\sigma}{\sigma}}\,\phi_{(l-1)}(r,\eta)
\label{13}
\end{equation}
where
\begin{equation}
\phi_{(l-1)}(r,\eta) = \,\sigma^{l-1}\,\underbrace{ {\partial_{r}}\,
{\frac{1}{\sigma}}\,\partial_{r}\,{\frac{1}
{\sigma}}\cdots
\partial_{r}\,\bigl(\frac{f+g}{\sigma}}_{l-1}\bigr).
\label{14}
\end{equation}
For $l=1$ we can write the right hand side of (\ref{12}) as
\begin{equation}
\phi_{1} = {\sigma} {\partial_{r}} {\frac{f_1+g_1}{\sigma}}.
\label{15}
\end{equation}
This satisfies our equation (\ref{9a}), as can be easily checked.
Now let $\phi_{l}$ be a  solution of:
\begin{equation}
(-{\partial_{\eta }}^2 + {\partial_{r }}^2) {\phi_{l}} =
{ l(l+1)\over \sigma^2}{\phi_{l}}.
\label{16}
\end{equation}
We claim that ${\phi_{(l+1)}}$ defined as in (\ref{13}) in
terms of ${\phi}_{l}$ satisfies the equation (\ref{16}) with
$l+1$ being put in place of $l$. That is easily proved by a direct
calculation using the relation (\ref{13})),
which shows that indeed  $(-{\partial_{\eta }}^2 + {\partial_{r }}^2)
{\phi_{(l+1)}} = {(l+1)(l+2)\over \sigma^2}{\phi_{(l+1)}}.$

Let us remark that   explicit  solutions of Eq. (\ref{9a}) are
known in the literature \cite{Friedlander}.  Eq. (\ref{9a})
has the same form as in the Minkowski spacetime (strictly saying, it
is so when $k=0$. If $k=1,-1$, Eq. (\ref{9a}) can be cast into the
Minkowskian form by  a suitable change of variables   (see for instance
\cite{Faraoni}, \cite{Noolan} and \cite{Bardeen}). The specific form of
the solution, presented in Eq. (\ref{12}), can be  new.

We would like to point out that the pair $(f,g)$ should be specified
independently for each multipole $\phi_l$; the related pair of functions
will be labelled by the multipole number $l$. The   initial data  for  the
wave equation (\ref{9a})  are  ${\phi_{l}}$, $\dot{\phi_{l}}$.
In the next section we show  that  those initial data suffice to determine
the pair $(f_l,g_l)$, which in turn determine solutions of Eq.(\ref{9a}) in
the whole spacetime.

\section{Generating functions and initial data}

Below we show how, having initial data one can construct the generating
functions $f$ and $g$ that appear in  the formula (\ref{12}).

Let   initial conditions be
\begin{eqnarray}
\phi_{l}(r,\eta)_l{\mid}_{\eta=0}\equiv{\phi_{l}(r,0)}&~~~~~{\partial_{\eta}}\,
 \phi_{l}(r,\eta)_l{\mid}_{\eta=0}
\equiv{{\dot{\phi}}_{l}(r,0)}
\label{17}
\end{eqnarray}
Assume that a support of initial data  is compact and  contained within
$(a,b)$. Integrating  $l$-times both sides of (\ref{12}), one arrives at
\begin{equation}
\bigl(g_{l}+f_{l}\bigr){\mid}_{\eta=0}=H_l(r) + M_l(r),
\label{18}
\end{equation}
where   quantities $H_l$ and    $M_l$ are linear combinations of multiple
integrals
\begin{eqnarray}
M_l&=& \hat M_0\sigma (r) +\sum_{s=1}^{l-1}   \hat M_s
 {\sigma}(r) \int_a^rdr_{1}  {\sigma}(r_1 )\nonumber \\
 && {\int_{a}^{r_{1}}dr_{2}
 {\sigma}(r_{2})  }  \cdots  \int_{a}^{r_{s-1}}dr_s  {\sigma}(r_s),
\nonumber\\
&&H_l(r)\,=\, { {\sigma}(r)} {\int_{a}^{r}dr_1 { {\sigma}
(r_{1})} }  {\int_{a}^{r_1}dr_2 { {\sigma}(r_{2})} }   \nonumber\\
&&\cdots{}
{\int_{a}^{r_{(l-1)}}   \frac{\phi_{l}(r,0)}{
{ {\sigma}^{l}(r_{l})}  }dr_l}.
 \nonumber\\
\label{19}
\end{eqnarray}
Here $\hat M$'s  are constants of integration.

From (\ref{12}), taking into account (\ref{1}) we  have
\begin{equation}
{{\dot{\phi }}_{l}(r,0)}\,=\sigma^l\underbrace{ {\partial_{r}}
{\frac{1}{\sigma }} {\partial_{r}}{\frac{1}{\sigma }}\cdots{\partial_{r}}
{\frac{\partial_{r}(g_{l}-f_{l})}{\sigma }}}_{l}.
\label{20}
\end{equation}
Therefore   performing an adequate number of integrations, one arrives at
\begin{equation}
\bigl(f_{l}\,-\,g_{l}\bigr){\mid}_{\eta=0}\,=-I_l-N_l,
\label{21}
\end{equation}
where
\begin{eqnarray}
I_l(r)&=&{\int_{a}^{r}dr_{1}{ {\sigma}(r_{1})}}   {\int_{a}^{r_{1}}
dr_{2}{ {\sigma}(r_{2})}} \nonumber\\
&&
\cdots
{\int_{a}^{r_{(l-1)}}dr_{l}{ {\sigma}(r_{l})}}  {\int_{a}^{r_{l}}
\frac{\dot{\phi}_{l}(r,0)}{{ {\sigma}^{l}(r_{(l+1)})}} \,
dr_{(l+1)}}. \nonumber\\
N_l&=& \hat N_0+ \sum_{s=1}^{l} \hat N_s \int_{a}^{r}dr_{1}      {\sigma}(r_{1})
\nonumber \\
&& {\int_{a}^{r_{1}}dr_{2}  {\sigma}(r_{2}) }
 \cdots  \int_{a}^{r_{s-1}}dr_s \sigma (r_{s}) \nonumber\\
\label{22}
\end{eqnarray}
and $\hat N$'s  are  the   integration constants.
The substraction of (\ref{21}) from  (\ref{18})  yields
\begin{eqnarray}
&& g_{l}{\mid}_{\eta=0}\;=\;\frac{1}{2}\bigl( H_l(r)+
I_l(r)\bigr) +
  {\frac{1}{2} } M_l(r)+  {\frac{1}{2} }N_l(r)
\label{23}
\end{eqnarray}
Similary, adding (\ref{21}) to (\ref{18})  gives
\begin{eqnarray}
&& f_{l}{\mid}_{\eta=0}\;=\;\frac{1}{2}\bigl( H_l(r,0) -
I_l(r,0)\bigr)
+{1\over 2}M_l - {\frac{1}{2} }N_l   .
\label{24}
\end{eqnarray}
Only quantities $H_l$ and $I_l$ carry information depending directly on
initial data. In contrast, let us stress out that the constants $\hat M$'s and
$\hat N$'s  are free -- hence $M_l$ and $N_l$ are   not determined
by initial data. Therefore the generating functions are not specified
uniquely by initial data. On the other hand, inverting this argument one can
say, that this nonuniqueness does  not influence   initial data. Since a
solution $\phi_{l}$ of Eq. (\ref{9a}) is uniquely determined by its initial
data, one infers that properties of the evolving waves do not depend on
specific values of the constants. Therefore we are free to adopt such values
that are convenient to us, and this fact   justifies the choice that
will be made at the end of this section.

The quantities $M_l$ and    $N_l$ are linear combinations of multiple
integrals. Using the additivity of
the integrals and splitting $\int_a^r=\int_a^b +\int_b^r$, one obtains
\begin{eqnarray}
M_l&=& \tilde M_0  \sigma (r)+\sum_{s=1}^{l-1}   \tilde M_s
 {\sigma}(r) \int_b^rdr_{1} {\sigma}(r_1 ){\int_b^{r_1}dr_2
 {\sigma}(r_{2})  }  \nonumber\\
 &&
 \cdots  \int_b^{r_{s-1}}dr_s
 {\sigma}(r_{s}),
\nonumber\\
N_l&=& \tilde N_0+ \sum_{s=1}^{l} \tilde N_s
\int_b^rdr_{1} {\sigma}(r_1)
\int_b^{r_1}dr_2  {\sigma}(r_2)   \nonumber\\
&&
\cdots  \int_b^{r_{s-1}}dr_s  {\sigma}(r_s) ,
\label{25}
\end{eqnarray}
with some constants $\tilde M$'s and $\tilde N$'s.  In a similar token
(but under additional condition $r>b$) one obtains
\begin{eqnarray}
H_l(r,0) & =& h_0  \sigma +\sum_{s=1}^{l-1}h_s
  \sigma (r) \int_b^rdr_1   \sigma (r_{1})
\int_b^{r_1}dr_2   \sigma (r_2) \nonumber\\
&&
\cdots{}
\int_b^{r_{(s-1)}}  dr_s  \sigma (r_s)
\nonumber \\
I_l(r,0) &=& i_0+ \sum_{s=1}^{l}i_s
    \int_b^rdr_1 \tilde \sigma (r_{1})
\int_b^{r_1}dr_2  \sigma (r_2) \nonumber\\
&&
\cdots{}
\int_b^{r_{(s-1)}} dr_s   \sigma (r_s).
\label{27}
\end{eqnarray}
Here $h$'s and $i$'s are fixed constants, that depend on initial data,
but whose specific form is not relevant. We would like to stress, that
here the condition $r>b$  is essential.

We  choose the   integration constants $\tilde M$'s, $\tilde N$'s
such that  the right hand sides of Eqs (\ref{23}) and (\ref{24})
become nullity for $r>b$.
Therefore  the generating functions $f_l, g_l$  vanish in
 the interval $(b, \infty )$ and become linear combinations $(\hat M\sigma ^+_-\hat N_0)/2$
 in $(0, a)$.

%
\section{The luminosity formula}

A lesson that can be drawn from the preceding section is this:
an outgoing wave pulse that is initially comprised within
$(r_E-\Delta , r_E)$ (as we assume from now on)
will remain inside $(r_E+\eta -\Delta , r_E+\eta )$.
A similar statement can be formulated about ingoing waves.
This provides an explicit proof of the  Huygens principle \cite{Hadamard}
in Friedman cosmological models \cite{remark1}.

In what follows we will restrict our attention  only to the outgoing
pulse of radiation having a width $\Delta $, and located initially
within $(r_E-\Delta , r_E)$.
The stress-energy tensor of the electromagnetic field
reads   $T_{\mu }^{\nu }= F_{\mu \gamma }F^{\nu \gamma }-
(1/4)g_{\mu }^{\nu }
F_{\gamma \delta }F^{\gamma \delta }$ and the time-like normal to a Cauchy
hypersurface is  $(n_{\mu }) =(-a, 0,0,0)$.
Define
\begin{eqnarray}
&&m(r,t) = -\int_{V(r)}dV aT_0^0 =\nonumber \\
&&2\pi \int_0^rdr \Sigma_{l=1}^{\infty }\Bigl( (\partial_{\eta }\phi_l)^2 +
(\partial_r\phi_l)^2+\frac{l(l + 1)}{\sigma^2}{\phi_l}^2
\Bigr) .
\label{32}
\end{eqnarray}
where $dV$ is the proper volume element and the volume  $V(r)$ extends
over the support of a radiation pulse that is enclosed inside the coordinate
sphere $S(r)$. $t$ is the cosmic time, related to the conformal time by
$dt=ad\eta $. One finds that
\begin{equation}
\partial_{t }m(r,t) ={4\pi \over a}\Sigma_{l=1}^{\infty }
 \partial_{\eta }\phi_l \partial_r\phi_l.
\label{33}
\end{equation}
The right hand side of Eq.(\ref{33}) can be interpreted
as  the "energy flux" through the comoving  coordinate sphere
$S(r)$. Assume that  $\Delta << \sigma (r_E)$ (we will refer later to this
condition as to the geometric optics approximation). Then it follows from
expression (\ref{12}) that one can approximate $\phi_l$ by its leading term,
$\partial_r^lf$, roughly speaking (see Appendix for a sketch of the precise
argument). Therefore one has approximate
equality valid {\it almost everywhere}
\begin{equation}
\partial_tm(r,t)  =
{-4\pi \over a} \Sigma_{l=1}^{\infty } (\partial_r^{l+1}f_l)^2.
\label{34}
\end{equation}
The total energy $m \equiv  \lim\limits_{r\rightarrow\infty}m(r,t)$
is strictly conserved, because  the energy flux vanishes outside
a coordinate sphere $S(r_0+\eta (t))$.
There is an interesting observation that can be made about the outgoing
electromagnetic radiation in the geometric optics limit.
If one  adopts our point of view in which  the emitter
is not a point   but rather a big sphere (say, a surface or an envelope
of a galaxy of a radius $r_E-\Delta $), then one can show that
$|T_{0r}|, T_{00}$ and $T_{rr}$ are much bigger that the remaining
components of the energy-momentum tensor.
In order to shorten  the mathematical formulae, we will operate with
quntities integrated over the coordinate sphere $r_E$. (In the
pointwise treatment one would have to deal  with expressions of the type
$\sum_{l,l'}Y_{l0}Y_{l'0} A_{ll'}$, where the symmetric tensor $A_{ll'}$
is given by  $A_{ll'}=(1/2)( \partial_r\phi_l\partial_\eta \phi_{l'}+
\partial_r\phi_{l'}\partial_\eta \phi_{l})$
or $A_{ll'}=(1/2)( \partial_r\phi_l\partial_r \phi_{l'}-\partial_{\eta }\phi_{l'}
\partial_{\eta }\phi_{l})$.) The dominant parts of
$\int_Sd^2S|T_{0r}|, \int_Sd^2ST_{00}$ and $\int_Sd^2ST_{rr}$ behave
like $\sum_l\partial_r\phi_l\partial_{\eta }\phi_l$; this can be
be approximated, in the limit of geometric optics, that is $\Delta << r_E$, by
$(1+0(\Delta /\sigma^{-1})) \sum_l(\partial_r^{l+1}f_l)^2$.
In contrast with that, the   surface integrals of the remaining components
of the energy momentum tensor are dominated by
$\sum_l\Biggl( (\partial_r\phi_l)^2-(\partial_{\eta }\phi_l)^2\Biggr) $. This
expression can be    approximated, in the same limit, by
$ 0(\Delta /\sigma^{-1}) \sum_l(\partial_r^{l+1}f_l)^2$.

Thus under the above conditions the electromagnetic radiation undergoes the
process of partial isotropisation and behaves like a   gas with nonisotropic pressure
(the tangent pressures $T_{\theta \theta }$ and $T_{\phi \phi }$ are negligible).
(That is an interesting fact in itself, which
in principle opens a way  for the explanation of the long standing riddle
concerning the apparent mass over-abundance of spiral galaxies \cite{Kutschera}.)

It was  claimed in \cite{Robertson} that all components of the energy-momentum
tensor different from $T_{r}^0, T_0^0$ and $T_r^r$ must vanish due to the isotropy of
the discussed cosmological models.  Robertson's statement is obviously wrong, but the
conclusion that $T_{r}^0, T_0^0$ and $T_r^r$ are the only components that matter
(in the limit of geometric optics, defined earlier), is correct.
Therefore the remaining part of the reasoning of \cite{Robertson}
can be left intact, modulo another  mistake mentioned in the introduction
(see also below).  Let us add here, for the convenience of an inspective reader,
that the quantity $ \Sigma_{l=1}^{\infty } (\partial_r^{l+1}f_{l})^2$
is equal (modulo a constant factor) to the quantity $V$ in the
paper of Robertson (see formula   (8) in \cite{Robertson}).

\subsection{The proper energy}
The comoving observer will detect the local electromagnetic
energy density $-T_{\mu \nu }n^{\mu }n^{\nu }=-T_0^0$. The total energy
of a wave pulse,
\begin{equation}
E(r,t )\equiv -\int_{V(r)}dVT_0^0,
\label{35}
\end{equation}
is related to the quantity $m$ by
\begin{equation}
E(r,t )=\frac{m(r,t)}{a(t)}.
\label{36}
\end{equation}
The volume $V(r)$ extends over an annulus $(r_E+\eta -\Delta ,r_E+\eta )$,
with the areal  radial  extension $a \Delta $  being of the order  of the
longest   wavelength present in the pulse.
The energy  flux through the comoving
coordinate sphere $S(r)$ reads
\begin{equation}
\partial_tE(r,t ) =  {-4\pi \over a^2}  (\partial_r^{l+1}(f_{l}))^2
-HE(r,t ).
\label{37}
\end{equation}
One finds, assuming (as in the preceding section) that $\Delta
<< \sigma (r_E)$, that
the energy  term $E(r,t)$ is approximated
by $4\pi \int_{r_E+\eta -\Delta }^{r_E+\eta }dr
\Sigma_{l=1}^{\infty }(\partial_r^{l+1}(f_{l}))^2/a\approx
(\Delta a)\Sigma_{l=1}^{\infty }(\partial_r^{l+1}(f_{l}))^2/a^2$.
It appears  that $H\approx 10^{-15}/km$ (remember that $c=G= 1$ ) -- that value
is prohibitively small, so that  for any realistic extension $\Delta $ of the radiation pulse
 the quantity $\Delta a H<<1$. Therefore the second term in (\ref{37}) is
 always negligible and   the energy flux  is well approximated   by
\begin{equation}
\partial_tE(r,t ) =  {-4\pi \over a^2}  \Biggl( \partial_r^{l+1}f_{l}\Biggr)^2.
\label{38}
\end{equation}
Notice that the factor   $(\partial_rf_{l0})^2$ is constant along
the trajectory of the outgoing wave pulse.
In order to compare this with Robertson, let us observe that  $\partial_tE(r,t )$
is equal (neglecting the $H$-related term) to $4\pi r^2a^2 U$  where $U$
is given by formula (2.10) in \cite{Robertson}.

\subsection{The luminosity distance}

We define $(r_E, t_E)$ and $(r_O, t_O)$ as
coordinates of the emitter and the observer, respectively
One easily finds, using (\ref{36}) and the constancy of $(\partial_rf_{l0})^2$,
that
\begin{eqnarray}
 \partial_tE(r,t )|_{t_O}&=&{a^2(t_E)\over a^2(t_O)} \partial_tE(r,t )\ |_{t_E} =\nonumber\\
&&{1\over (1+z)^2} \partial_tE(r,t ) |_{t_E}.
\label{39}
\end{eqnarray}
Notice that  $ \partial_tE(r,t ) |_{t_E}$ is the luminosity of the emitted radiation.
The apparent luminosity is given by the ratio $ \partial_tE(r,t ) |_{t_O}/(4\pi R_O^2)$.
From the definition of the luminosity distance follows now immediately
\begin{equation}
D=(1+z)R_0,
\label{40}
\end{equation}
in accordance with the formula  (\ref{0}).

\section{Appendix}

Let $f\in C^{\infty }$ have a compact support $(a,b)$ such that $\Delta =b-a<<a$.
It is easy to see that $ ||\partial^lf  /\sigma ||_{L_2(a,r)} <<||\partial^{l+1}f   ||_{L_2(a,r)}$,
for any interval $(a,r)\in (a,b)$. Therefore the amount $m_{(a,r)}$  of  radiation within the
annulus $(a,r)$ is well approximated by $||\partial^{l+1}f   ||_{L_2(a,r)}^2$
and the pointwise equality of Eq. (\ref{34}) holds true almost everywhere.


\begin{references}
\bibitem{1} The areal radius is not directly measurable. Eq. (\ref{1})
implies that $D=(1+z)^2r_A$, where $r_A$ is the angular distance, which may be
determined directly.
\end{references}
\end{document}